\documentclass[12pt]{article}
\usepackage{amsmath}

\usepackage{natbib}
\setcitestyle{maxcitenames=6}
\usepackage{graphicx}
\usepackage{url}
\usepackage{float}
\usepackage{color}
\usepackage{caption}
\usepackage{subcaption}
\usepackage{array}

\newcommand{\blind}{0}

\begin{document}

\def\spacingset#1{\renewcommand{\baselinestretch}%
{#1}\small\normalsize} \spacingset{1}

%%%%%%%%%%%%%%%%%%%%%%%%%%%%%%%%%%%%%%%%%%%%%%%%%%%%%%%%%%%%%%%%%%%%%%%%%%%%%%

\if0\blind
{
  \title{\bf 
  survivalContour: Visualizing predicted survival %for continuous covariates
  via colored contour plots}
  \author{Yushu Shi,  \\
  %\thanks{
   % The authors gratefully acknowledge funding from NSF 2310955 [Y.S.], 
   % NIH/NCI CCSG P30CA016672, MD Anderson Moon Shot Programs, Prostate Cancer SPORE P50CA140388, CCTS 5UL1TR000371, and CPRIT RP160693 [K.A.D.], 
   % NIH R01 HL158796 and NIH/NCI CCSG P30CA016672 [C.B.P.], and  
   % NIH R01 HL124112, NIH R01 HL158796, and CPRIT  RR160089 [R.R.J.]}\hspace{.2cm}\\
    {\small Department of Population Health Sciences, Weill Cornell Medicine}\\[5pt]
    Liangliang Zhang,\\
    {\small Department of Population and Quantitative Health Sciences,}\\  {\small Case Western Reserve University} \\[5pt]
    Kim-Anh Do,\\
     {\small Department of Biostatistics,}\\  {\small The University of Texas MD Anderson Cancer Center}\\[5pt]
     Robert R.\ Jenq,\\
     {\small Department of Genomic Medicine,}\\  {\small The University of Texas MD Anderson Cancer Center}\\[5pt]
     and \\
     Christine B.\ Peterson\\
     {\small Department of Biostatistics,}\\  {\small The University of Texas MD Anderson Cancer Center}
    }
  \maketitle
} \fi

\if1\blind
{
  \bigskip
  \bigskip
  \bigskip
  \begin{center}
    {\LARGE\bf Title}
\end{center}
  \medskip
} \fi

\begin{abstract}
Advances in survival analysis have facilitated unprecedented flexibility in data modeling, yet there remains a lack of tools for graphically illustrating the influence of continuous covariates on predicted survival outcomes. We propose the utilization of a colored contour plot to depict the predicted survival probabilities over time, and provide a Shiny app and R package as implementations of this tool. Our approach is capable of supporting conventional models, including the Cox and Fine-Gray models. However, its capability shines when coupled with cutting-edge machine learning models such as random survival forests and deep neural networks.
\end{abstract}

\noindent \textbf{Running title}: Survival contour plots

\noindent \textbf{Additional information}: Christine B. Peterson, The University of Texas MD Anderson Cancer Center, 7007 Bertner Ave, Houston, TX, 77030. \url{cbpeterson@mdanderson.org}\\
\textsl{The authors declare no potential conflicts of interest.}
%\noindent%
%{\it Keywords:} deep learning, interactive graphics, visualization, R package, Shiny app, survival prediction
%\vfill

\newpage
\spacingset{1.45} % DON'T change the spacing!
\section{Introduction}
\label{sec:intro}

Over the past few decades, there have been major advances in survival modeling, including improvements in approaches for handling competing risks and the development of machine learning methods for survival prediction.
However, there is a lack of appropriate visualization tools for
translating these cutting-edge models into effective graphical representations that can illustrate the relationships between continuous predictors and survival probabilities in an intuitive manner. Even for classical models, such as the Cox or Fine-Gray methods, there is a gap between the model outputs and the presentation of results.

\begin{figure}[H]

\textbf{Binary representations with continuous predictor split at the median}

    \begin{subfigure}[t]{0.49\textwidth}
    \includegraphics[width=0.92\linewidth,trim={0 0 0 0.85cm}, clip]{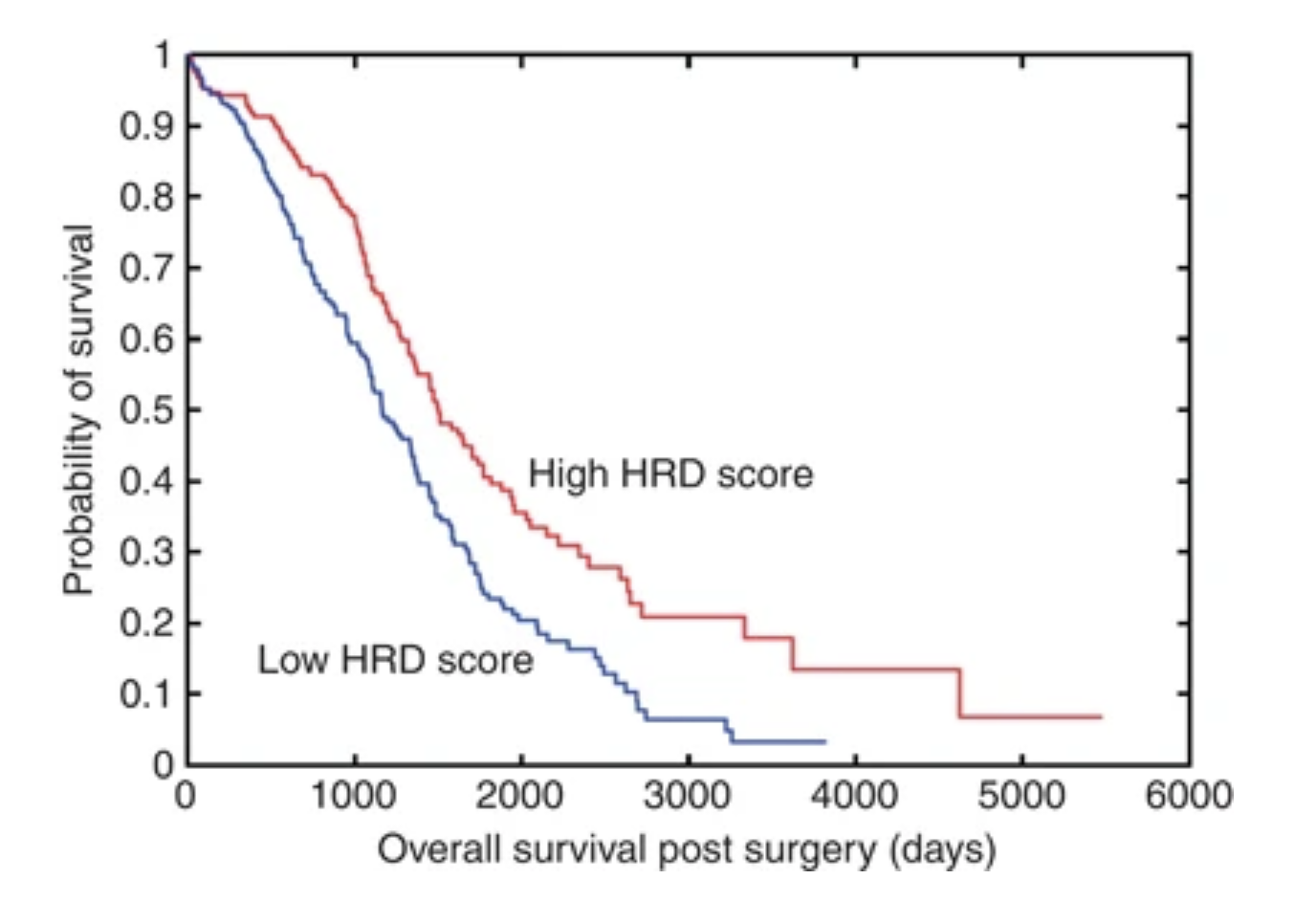}
    \caption{HRD score  \citep{AbkeTimm12}} 
    \label{fig:bin1}
    \end{subfigure} \hfill
    \begin{subfigure}[t]{0.49\textwidth}
    \centering
    \includegraphics[width=0.75\linewidth,trim={1cm 0.8cm 0 -0.5cm}, clip]{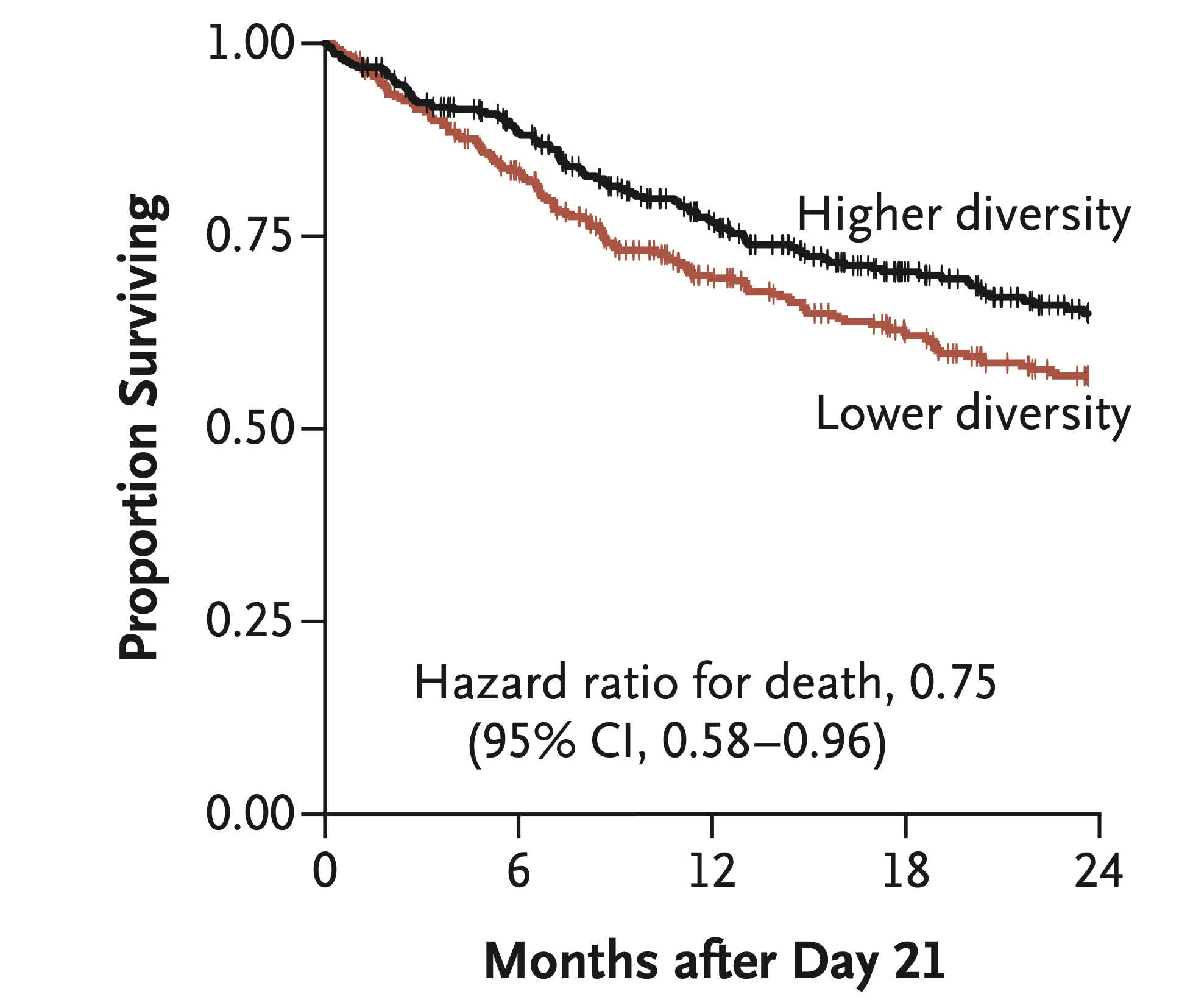}
     \caption{Microbiome diversity \citep{PeleGome20} }
     \label{fig:peled}
\end{subfigure}

\vskip.3cm
\textbf{Multi-group representations with split at extreme quantiles}

\begin{subfigure}[t]{0.49\textwidth}
    \includegraphics[width=0.9\linewidth,trim={0 0.5cm 0 -0.5cm}, clip]{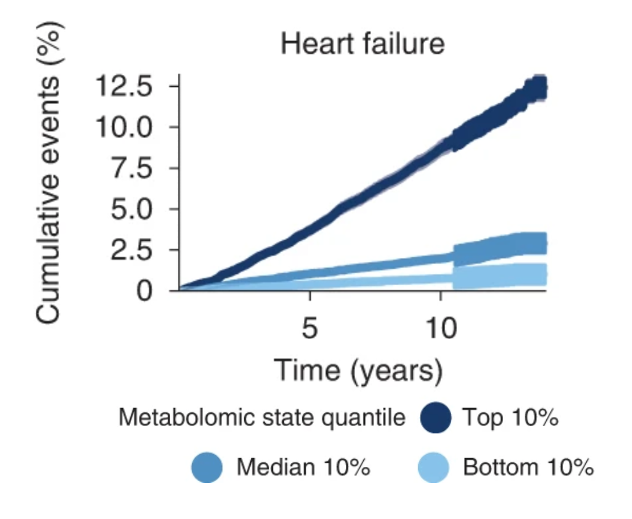}
    \caption{Metabolomic state  \citep{BuerStei22}}
    \label{fig:quantiles1}
    \end{subfigure} \hfill
    \begin{subfigure}[t]{0.49\textwidth}
    \centering
    \includegraphics[width=1.1\linewidth,trim={1.2cm 0.5cm 0 1.2cm}, clip]{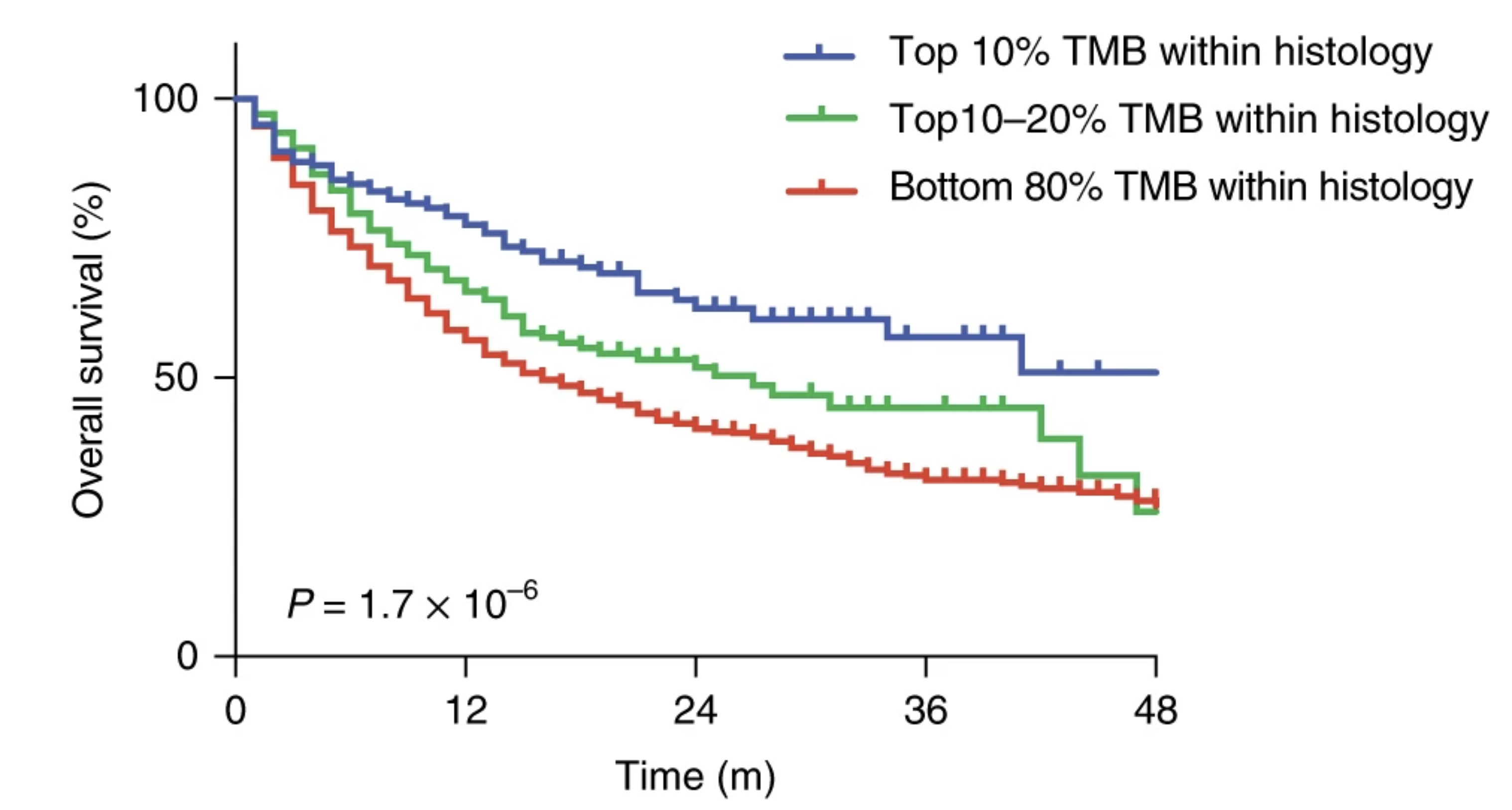}
     \caption{TMB \citep{samstein2019}}
     \label{fig:quantiles2}
\end{subfigure}

\vskip.3cm
\textbf{Spaghetti plots with predicted survival curves for individual subjects}
    
\begin{subfigure}[t]{0.49\textwidth}
    \includegraphics[width=0.93\linewidth,trim={0 0.2cm 0 0.2cm}, clip]{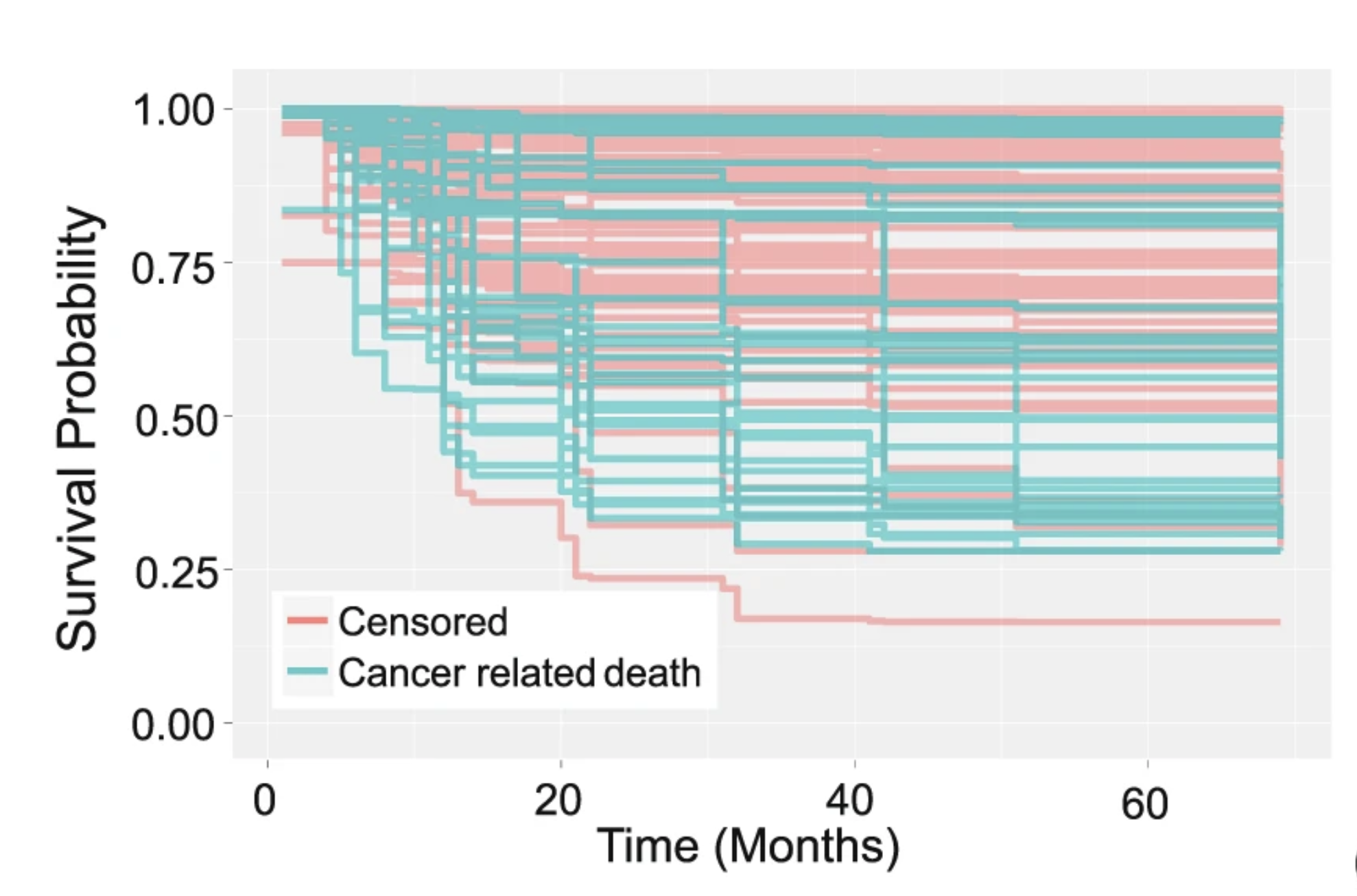}
    \caption{Random survival forests predictions colored by group  \citep{KimLee19}}
    \label{fig:spag1}
    \end{subfigure} \hfill
    \begin{subfigure}[t]{0.49\textwidth}
    \centering
    \includegraphics[width=0.90\linewidth,trim={0.8cm 0.2cm 0 0cm}, clip]{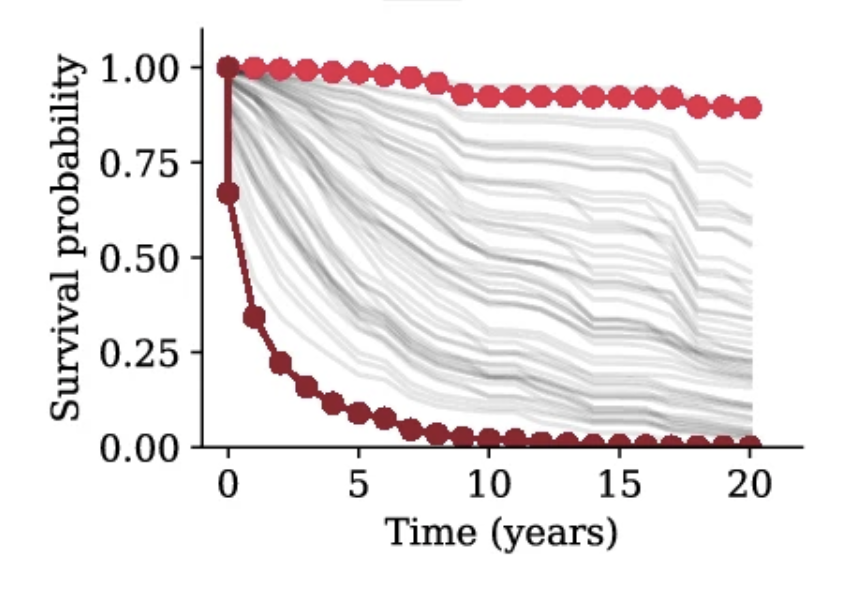}
     \caption{Deep learning predictions with extreme patients highlighted  \citep{ValeSilva2021}}
     \label{fig:spag2}
\end{subfigure}
\caption{Examples of existing visualization methods}

\end{figure}

A typical visualization approach in current practice is to split the covariate of interest at its median, and present Kaplan-Meier curves for subjects with low vs.\ high values. Figures \ref{fig:bin1} and \ref{fig:peled} display two examples from the biomedical literature where Kaplan-Meier curves were obtained from binarizing a continuous predictor at the median value. 
This approach has significant shortcomings, as precise information from the continuous predictor gets lost during binarization \citep{AltmRoys06, FernMala19, MacCZhan02}. In addition, the $p$-value for the coefficient in the Cox model
using the continuous version of the predictor does not necessarily match the $p$-value that one would obtain from a log-rank test comparing
the two Kaplan-Meier curves. This mismatch creates a divergence between the analysis and the visual presentation. This inconsistency is clear in Figure \ref{fig:peled}, where a hazard ratio derived from a Cox model is displayed alongside a two-group representation of patient survival.

To highlight interesting trends for extremes of the predictor distribution, narrower categories may be constructed using quantiles of the predictor. However, this step introduces an additional element of subjective data exploration in choosing how to represent the association between the continuous predictor and survival. Figures \ref{fig:quantiles1} and \ref{fig:quantiles2} depict examples from the literature where tail quantiles, such as the bottom 10\% or top 10\%, were identified to highlight differences in the survival curves for patients with extreme predictor values. 

The lack of flexible and informative visualization approaches becomes even more prominent when considering advanced machine learning-based methods, which relax the assumption of proportional hazards or accelerated failure times. Foundational work on the application of machine learning methods to survival prediction includes the random survival forests \citep{IshwKoga08}.
More recently, there has been a flurry of proposals on the use of deep learning for survival prediction, including DeepHit \citep{LeeZame18}, DeepSurv \citep{KatzShah18}, Nnet-survival \citep{GensNara19}, and pycox \citep{HavaBorg19, HavaOrnu21}. The challenge of presenting survival predictions generated by these approaches is particularly pressing given the complex non-linear relationships they encode.

To visualize model outputs, spaghetti plots can be used to display predicted survival or cumulative incidence curves for individual patients. In spaghetti plots, which have traditionally been applied to depict trends in longitudinal data, a distinct trajectory, or ``noodle",  is plotted for each subject \citep{Szabo2009}. Figures \ref{fig:spag1} and \ref{fig:spag2} highlight the use of spaghetti plots to display predicted survival from random survival forests and deep learning models. %ShamShen21
Although the curve for an individual subject may be informative, when applied to a patient cohort spaghetti plots tend to result in a tangle of overlapping curves without shedding light on the link between covariate values and predicted survival.

Alternative approaches to visualization have been proposed in the statistical literature. In particular, the idea of visualizing survival outcomes through a contour plot appeared in \cite{LumlHeag00}. However, their method was implemented in XLisp-Stat, which is not widely used today, and does not reflect developments in the field of survival analysis over the past two decades. % involving competing risks, interval censored data, or neural-network based models. 
Currently available tools for survival visualization include the \texttt{survminer} package \citep{survminer}, which can produce publication-ready plots and risk tables, but still focuses on the display of Kaplan-Meier and cumulative incidence curves. Most recently, \cite{denz2023} proposed the use of a survival area plot for depicting the effect of a predictor within the causal inference framework. However, there remains a lack of user-friendly visualization tools for presenting the results of popular survival models, especially machine learning based models. To address this gap, we have designed a Shiny app for interactive visualization and also provide an R package for more advanced users. Both tools enable the production of visually effective colored contour plots and offer flexible options for handling competing risks and interval censoring.
Our software captures a broad set of models for survival association and offers more advanced users seamless integration to deep learning models.

% To address the gap in user-friendly visualization tools, we propose the use of colored contour plots to depict  predicted survival probabilities. We 

\begin{figure}[t]
	\includegraphics[width=\textwidth]{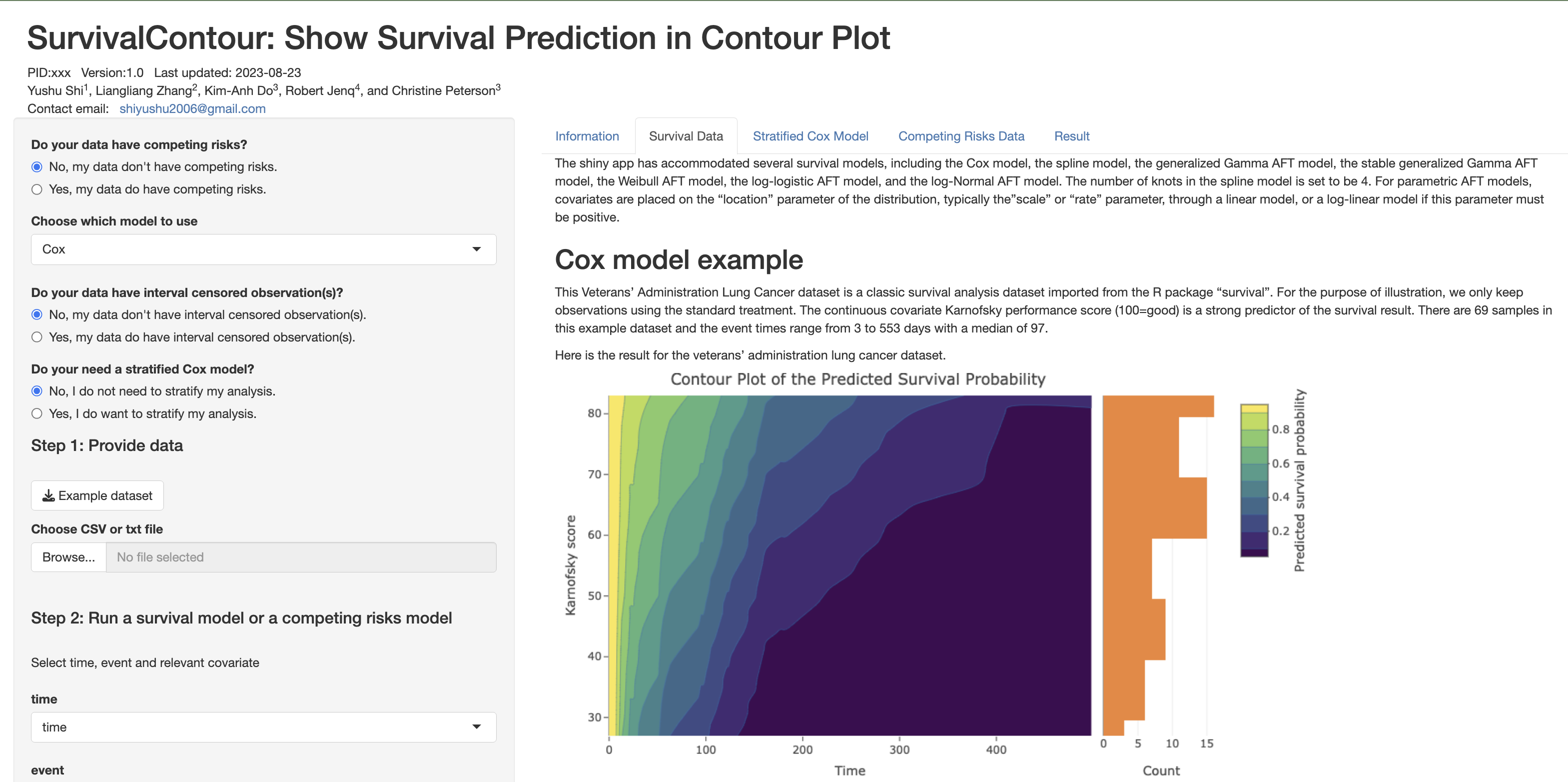}
	\caption{A snapshot of the survivalContour Shiny app}
	\label{snapshot}
	\end{figure}

\section{Materials and Methods}
\subsection{Shiny app}
\begin{figure}[th]
\centering
	\includegraphics[width=1.05\textwidth]{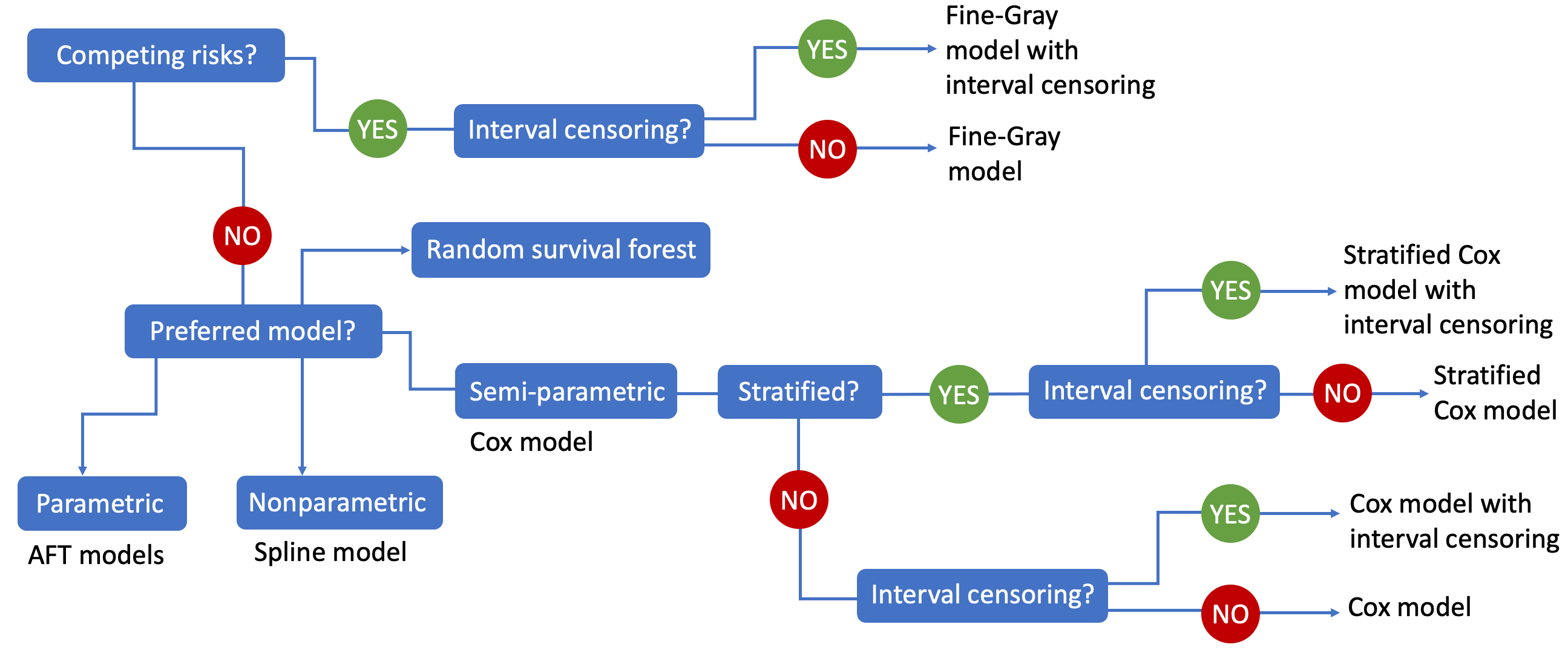}
    \caption{Decision tree implemented in the Shiny app }
    \label{fig:dec}
\end{figure}

We have designed a user-friendly Shiny app for automatically generating survival contour plots, available at \url{https://biostatistics.mdanderson.org/shinyapps/survivalContour/}.
A snapshot of the survivalContour Shiny app is shown above in Figure \ref{snapshot}.
In the Shiny app, we guide users to select an appropriate survival model for their data. There are options for settings with competing risks, where the event of interest may be precluded by a different event, and for interval-censored data, where the event time is only known to occur within a given window. Competing risks and interval censoring are both common in biomedical studies. For example, competing risks occur when a patient is no longer at risk for cancer progression because they have died due to an infection. Interval censoring occurs whenever there are fixed time points for screening or follow-up. For example, in a cancer screening study, if a scan reveals that the patient developed a tumor since their last visit, the true time-to-event lies somewhere during the intervening time window but is not exactly observed. For settings without competing risks, we offer parametric, nonparametric, semi-parametric, and machine learning based survival models, as these may be preferred for different applications. Parametric models, which assume a known distribution for the survival times, offer an advantage for settings where out-of-sample prediction is a priority \citep{Jack16}. Nonparametric survival models, such as spline-based methods, offer greater flexibility on the shape of the hazard \citep{RoystonParmar}. Semi-parametric methods, in particular the Cox model, which avoids assumptions on the baseline hazard function, are a mainstay in applied survival analysis. We offer both the classic and stratified Cox model, which allows the baseline hazard function to differ across levels of a covariate. Finally, machine learning-based methods offer the most flexible modeling of survival data. Although they do not allow for inference on the predictors, those methods can provide predicted survival probabilities for given covariates at specific times.
% We first
% ask whether their data set includes competing risks. For data sets without competing risks, users are provided modeling choices including non-parametric (spline), semi-parametric (Cox), and fully parametric models (AFT). For data sets with competing risks, users are directed to either the Fine-Gray model \cite{FineGray} or the Fine-Gray model with interval censoring.
These options are illustrated as a decision tree in Figure \ref{fig:dec}.

Once the desired model has been selected, we ask users to upload their data, designate the time and censoring indicator, and select the continuous predictor for visualization. We then produce a 2D contour plot: in this plot, the x-axis is time, the y-axis is the continuous predictor of interest, and the intensity of regions in the plot corresponds to the predicted survival probability from the fitted model. A histogram showing the distribution of the continuous predictor is displayed alongside the contour plot. 
 In addition to the major continuous covariate of interest for visualization, users can provide other covariates to be included as adjusters in the predictive model. The survival predictions presented in the contour plots are conditional on the covariate values: these may be specified to a desired level by the user, or taken as the default of the median value for continuous covariates and the most frequent value for categorical covariates. 

\subsection{R package}
In addition to the Shiny app, we provide an R package for more advanced users, which allows the display of predictions from deep neural network-based models. Because of the manual tuning involved with deep neural network-based models, we restrict the choice of these models to the R package only. The survivalContour R package can be downloaded from the GitHub page \url{https://github.com/YushuShi/survivalContour/}. We list the models that are compatible with our tools in Table \ref{supportTable}. 

% \begin{table}
% \begin{tabular}{llll}\hline
% 	 Model & Package & Object &Confidence interval\\
% 	 &&& in 3D plot\\\hline
% 	 Cox model& survival& coxph & Y \\\hline
% 	 Stratified Cox model& survival &coxph& Y\\\hline
% 	 Cox model with& mets &phreg& Y \\
% 	 interval censored data&&& \\\hline
% 	 Stratified Cox model with& mets &phreg& Y \\
% 	 interval censored data& && \\\hline
% 	 Parametric and spline models & flexsurv & flexsurvreg&Y \\\hline
% 	 Fine-Gray model & riskRegression & FRG&N \\
%      for competing risks data & && \\\hline
% 	 Fine-Gray model&intccr &ciregic& N \\
% 	 %for competing risks data& && \\
% 	 with interval censoring& &&  \\\hline
% 	coxtime, deepsurv, deephit&survivalmodels&pycox&N\\
% 	loghaz, pchazard& &&\\\hline
% 	\end{tabular}
% 		\caption{Models and packages survivalContour supports}
% 	\label{supportTable}
% 	\end{table}
 
\begin{table*}[ht]
 \setlength\extrarowheight{-1pt}
  \setlength{\baselineskip}{0.1\baselineskip}
\begin{tabular}{p{9cm}p{6cm}}\hline
	 Model & Package\\ \hline
	 Cox model and stratified Cox model \citep{Cox72, TherGram00} & survival \citep{Ther23}\\ \hline
	 Cox model and stratified Cox model for interval-censored data& mets \citep{ScheHols14, HolsSche16}  \\ \hline
	 Parametric and spline models \citep{RoystonParmar} & flexsurv \citep{Jack16}\\ \hline
	 Fine-Gray model for competing risks data \citep{FineGray99} & riskRegression \citep{GerdKatt21, GerdSeba23}\\ \hline
	 Fine-Gray model with interval censoring \citep{BakoYu17} &intccr \citep{ParkBoka19} \\ \hline
  random survival forests \citep{IshwKoga08} &randomForestSRC \citep{IshwKoga23} \\ \hline
	Cox-Time \citep{HavaBorg19} &survivalmodels \citep{Sona22}\\
 DeepSurv \citep{KatzShah18}&\\
  DeepHit \citep{LeeZame18}&\\
	Nnet-Survival \citep{GensNara19}& \\
 PC-Hazard \citep{HavaOrnu21}& \\\hline
	\end{tabular}
		\caption{Models and packages supported by survivalContour}
	\label{supportTable}
	\end{table*}

 \subsection{3D plots}
Both the Shiny app and the R package support visualization of the survival contours in three dimensions. This provides a more comprehensive understanding of predicted survival, since a two-dimensional (2D) colored plot essentially represents a top-down view of the three-dimensional (3D) structure. Additionally, the three dimensional representation allows for the illustration of confidence intervals as semi-transparent layers.

\section{Results}
In this section, we illustrate the capacity of our Shiny app and R package to present predictions from standard Cox models, stratified Cox models, competing risks regression, random survival forests, and deep learning methods.

\subsection{Standard Cox model} \label{sec:cox}
To highlight the utility of the Shiny app for presenting the results from a standard Cox model, we reanalyzed the data presented in Figure \ref{fig:peled}, which was originally published in \cite{PeleGome20}. This study established that lower microbiome diversity at baseline was associated with increased risk of mortality following allogeneic hematopoietic-cell transplant across multiple independent patient cohorts. Here, we obtained the data from Cohort 1, which was composed of patients treated at Memorial Sloan Kettering Cancer Center. Microbiome diversity was computed using the inverse Simpson index.
 In the presentation of the data shown in Figure \ref{fig:peled}, patients were separated into low vs.\ high diversity groups based on the median value of diversity for samples collected between Days 7 and 21. The Kaplan-Meier curves for each group were presented alongside the hazard ratio for death from a Cox model with the log10 of diversity as the primary predictor of interest.
 
\begin{figure}[H]
\begin{subfigure}[c]{\textwidth}
\centering
    \includegraphics[width = \linewidth]{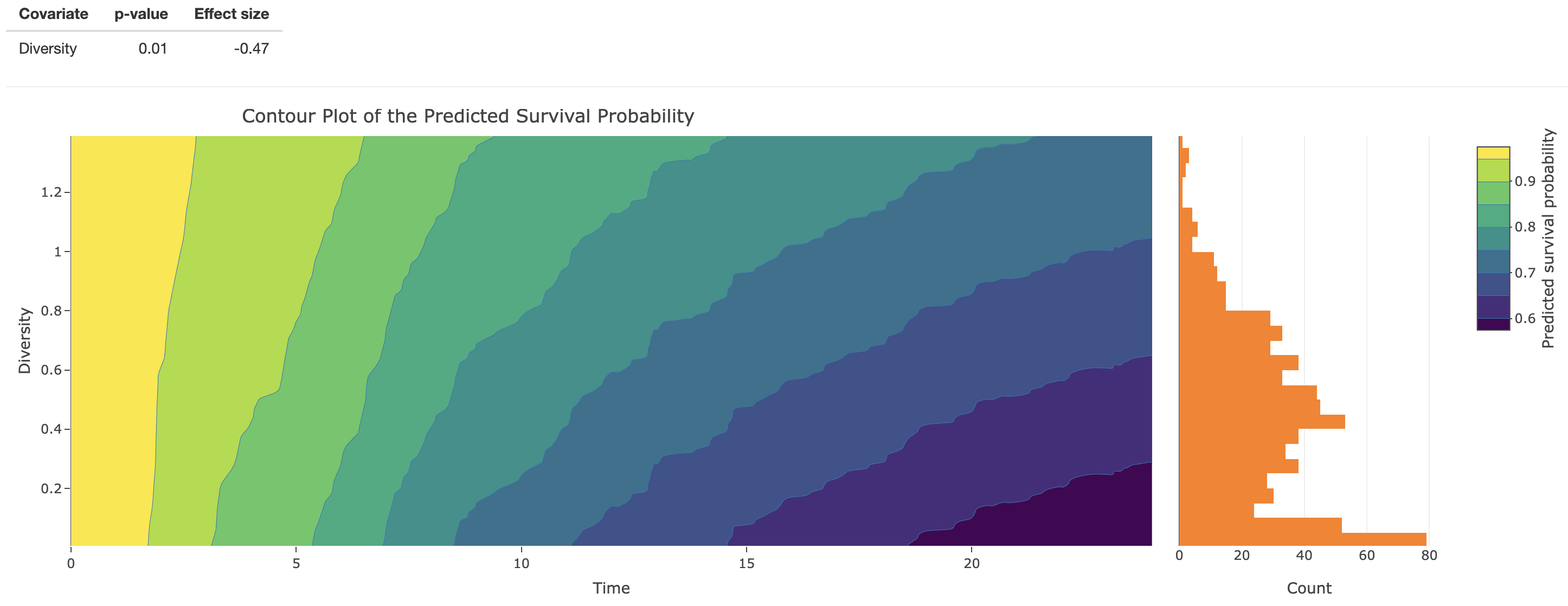}
\caption{Contour plot. Diversity corresponds to microbiome diversity on the log10 scale.}
\label{fig:peled_new1}
\end{subfigure}
\begin{subfigure}[c]{\textwidth}
    \includegraphics[width = 0.8\linewidth]{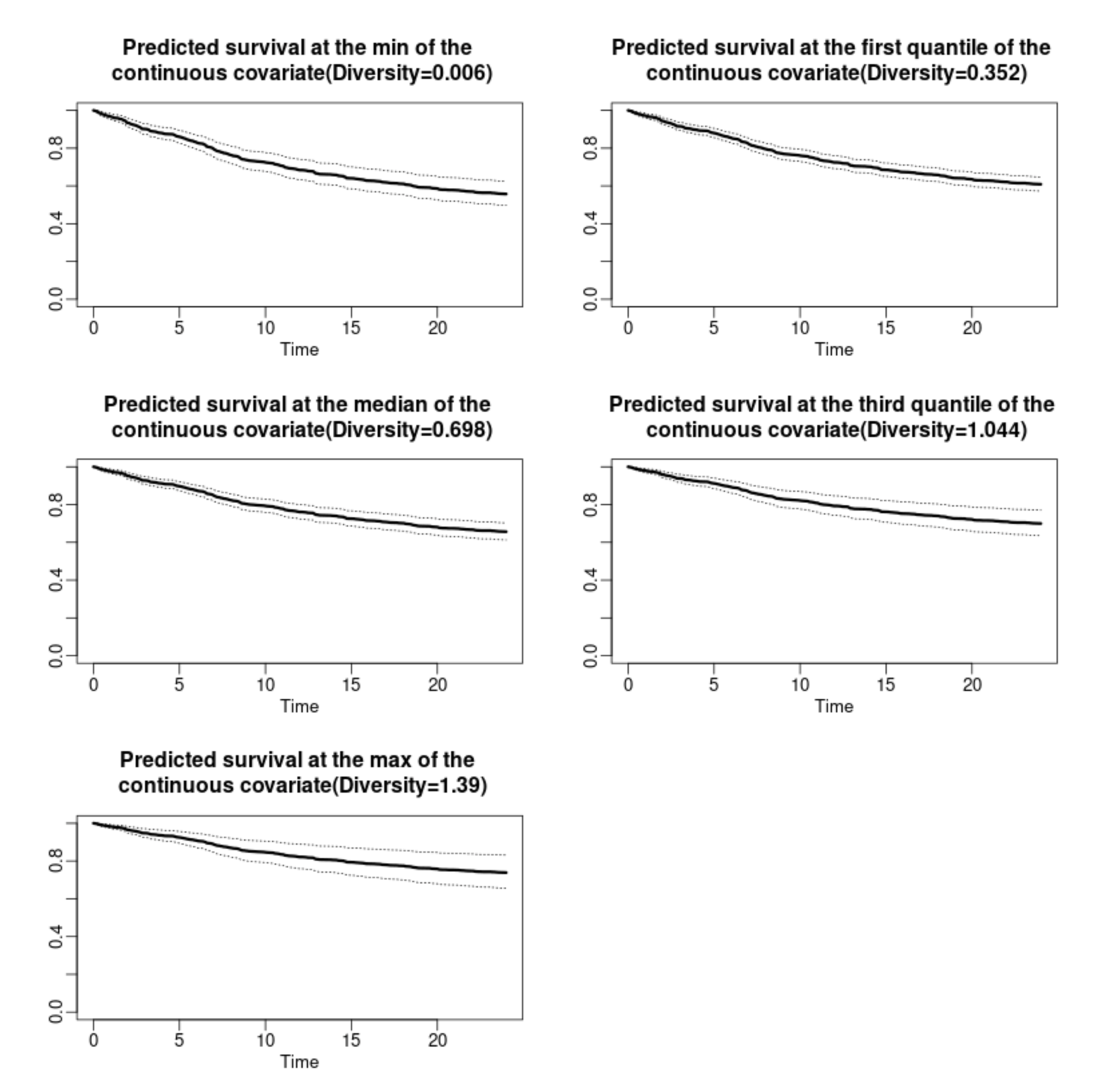}
\caption{Predicted survival curves. Diversity corresponds to baseline microbiome diversity on the log10 scale.}
\label{fig:peled_new2}
\end{subfigure}
\caption{Plots created by survivalContour Shiny app using \cite{PeleGome20}'s data}
\end{figure}

 A richer representation of this association is presented in Figure \ref{fig:peled_new1}, which depicts a contour plot of the predicted survival from the Cox model. 
 In this plot, the x-axis represents time in months after Day 21, the y-axis represents microbiome diversity on the log10 scale, and the color intensity represents the predicted survival probability. As an example of how to interpret the plot, at 10 months, a subject with microbiome diversity $< 0.2$ on the log10 scale would have a predicted survival probability in the range 70-75\%, while a subject with a diversity value of 1.2 on the log10 scale would have a predicted survival of 80-85\%. This plot corresponds more closely to the analysis performed than the Kaplan-Meier curves and offers an intuitive representation of the strength of association between the predictor and the outcome.  The orange histogram at right provides  information on the distribution of the predictor values as well as the number of subjects included in the analysis (via the bar heights).  Here, the bar heights provide a key insight that very low diversity values (which confer worse expected survival) are in fact quite common in the observed data for this cohort, as reflected by the histogram mode occurring at the lowest bin.

In Figure \ref{fig:peled_new2}, we show the predicted survival curves for 5 quantiles of the predictor values. These plots are automatically generated by the app, and provide a snapshot of the predicted survival at evenly spaced quantile values. Importantly, these plots include confidence intervals to highlight the degree of statistical uncertainty in the predictions.

Here, we presented the results from a univariate Cox model for the sake of simplicity. However, the survivalContour app allows for the inclusion of covariates in the model.

\subsection{Stratified Cox model}
\label{sec:stratified}
To demonstrate the capabilities of the survivalContour Shiny app in tandem with the stratified Cox model, we provide an illustrative example using the Veterans' Administration Lung Cancer study \citep{KalbPren11}, available in the \texttt{survival} R package. The data set includes covariate data and survival outcomes for 137 subjects. Our focus is on elucidating the influence of the Karnofsky performance score \citep{karn} on survival times. Since this relationship varies by cancer subtype, we fit a stratified Cox model, with strata defined by the four subtypes in the dataset: adenocarcinoma,   large cell carcinoma, small cell lung cancer, and squamous cell carcinoma.
We adjusted for the additional covariates age, prior therapy status,  treatment arm, and months from diagnosis to randomization.
The predicted survival probabilities are shown in Figure \ref{stratified}, where the projected survival for each cancer type is presented in a subpanel. Higher Karnofsky scores are associated with improved survival predictions across all four subtypes: however, the strength of this association and overall hazard rates vary by cancer subtype.

\textbf{\begin{figure}[H]
\centering
	\includegraphics[width=1.03\textwidth]{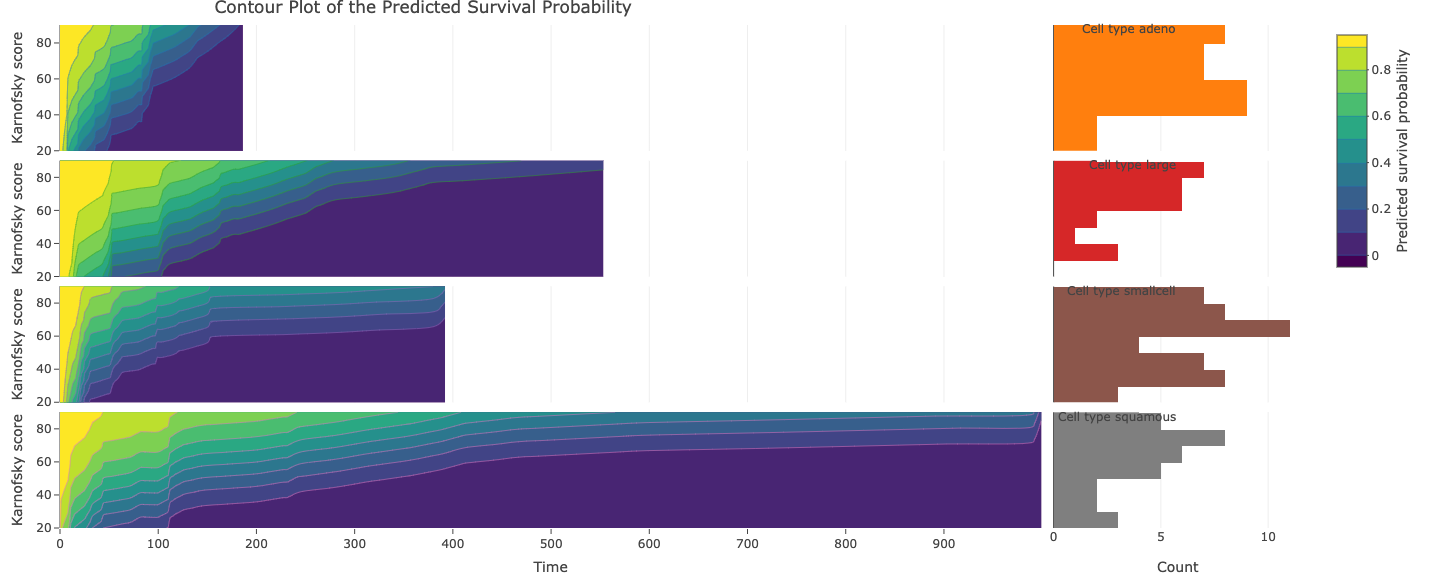}
	\caption{Plot generated by the survivalContour R package for stratified analysis. Each subpanel corresponds to a lung cancer subtype: adenocarcinoma (orange), large cell carcinoma (red), small cell carcinoma (brown), and squamous cell carcinoma (gray).}
	\label{stratified}
\end{figure}
}
\subsection{Competing risks}
\label{sec:cmprsk}
The survivalContour Shiny app, along with its accompanying R package, can handle data with competing risks. To demonstrate this capacity, we use the Paquid dataset, which was collected as part of a study on brain ageing \citep{DartGagn92} and is included in the \texttt{riskRegression} 
package \citep{GerdSeba23}. The data set includes a total of 2561 participants, with observed event times for the onset of dementia for 449 subjects and for death without dementia for 634 subjects.
We focused on predicting the onset of dementia  while considering death without dementia as a competing risk. 

As our primary predictor variable of interest, we considered 
the Digit Symbol Substitution Score Test (DSST), which explores attention and psychomotor speed. Concurrently, we adjust for the Mini-Mental State Examination score (MMSE), which is a measure of overall cognitive performance. Employing the Fine and Gray model, we illustrate the influence of DSST in Figure \ref{FG} via both 2D and 3D contour representations. Here, lower scores on the DSST are associated with an increased probability of developing dementia over time, accounting for the fact that subjects that died without dementia were removed from the risk pool.

\textbf{\begin{figure}[H]
\centering
	\includegraphics[width=0.6\textwidth]{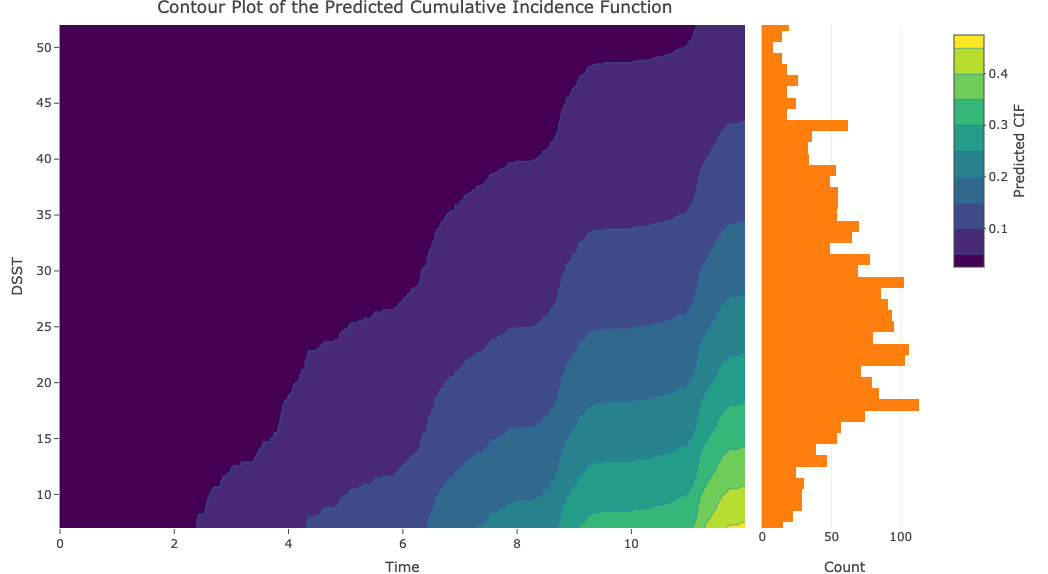}
 	\includegraphics[width=0.37\textwidth]{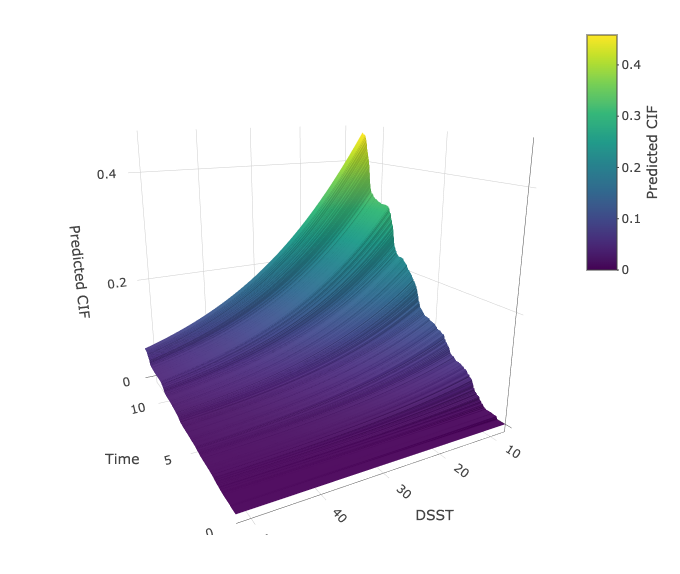}
	\caption{2D plot (left) and 3D plot (right) generated by the survivalContour R package for the Paquid competing risks data. }
	\label{FG}
\end{figure}
}

\subsection{Machine learning methods}
\label{sec:ml}
The proposed contour plot visualization shines when integrated with advanced machine learning models like random survival forests and deep neural networks. %In this section, we present two illustrative examples. 
%In both examples, the connection between the predictor and the predicted survival is non-monotonic, posing a challenge for conventional survival models.
In this section, we rely on data collected by the SUPPORT III study \citep{KnauHarr95}, which aimed to characterize survival outcomes of seriously ill hospitalized adults. This dataset, which
 contains 8873 observations, was also analyzed in the 
  paper introducing the DeepSurv method \citep{KatzShah18}. % and in the pycox Python package \citep{Kvam23}.
 In this case study, we are interested in depicting the influence of respiratory rate on survival. Additional covariates in the data set include age, sex, race, the number of comorbidities, the presence of cancer, diabetes, and dementia, mean arterial blood pressure, temperature, white blood count, heart rate, and serum creatine. 

\begin{figure}[H]

\centering
\begin{subfigure}[c]{\textwidth}
	\includegraphics[width=0.49\textwidth]{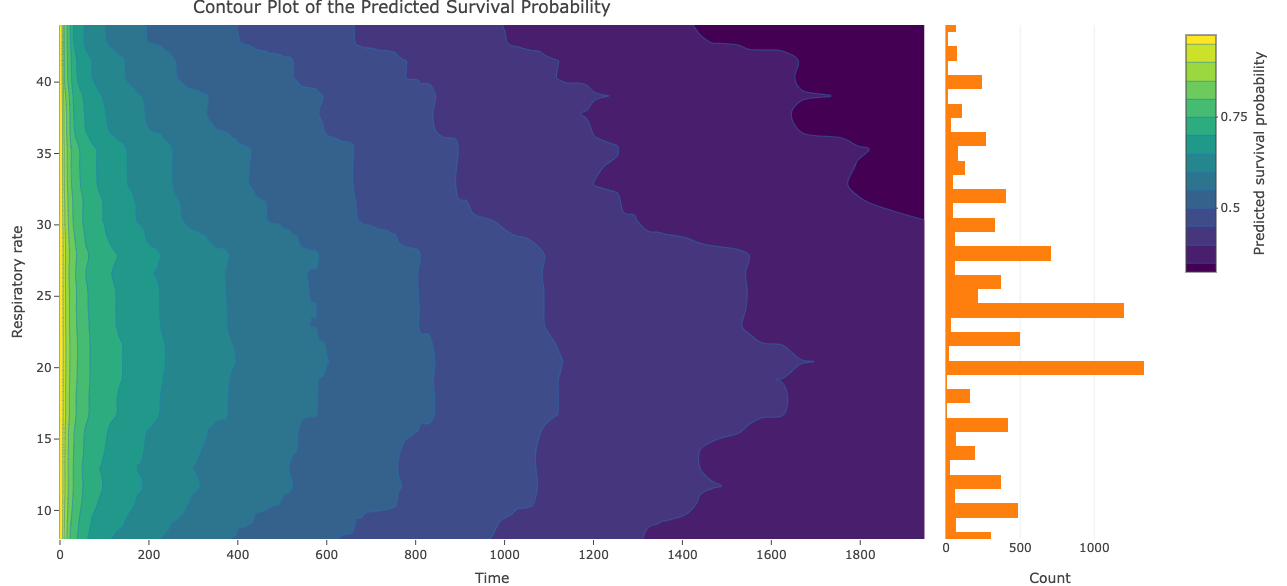}
 	\includegraphics[width=0.49\textwidth]{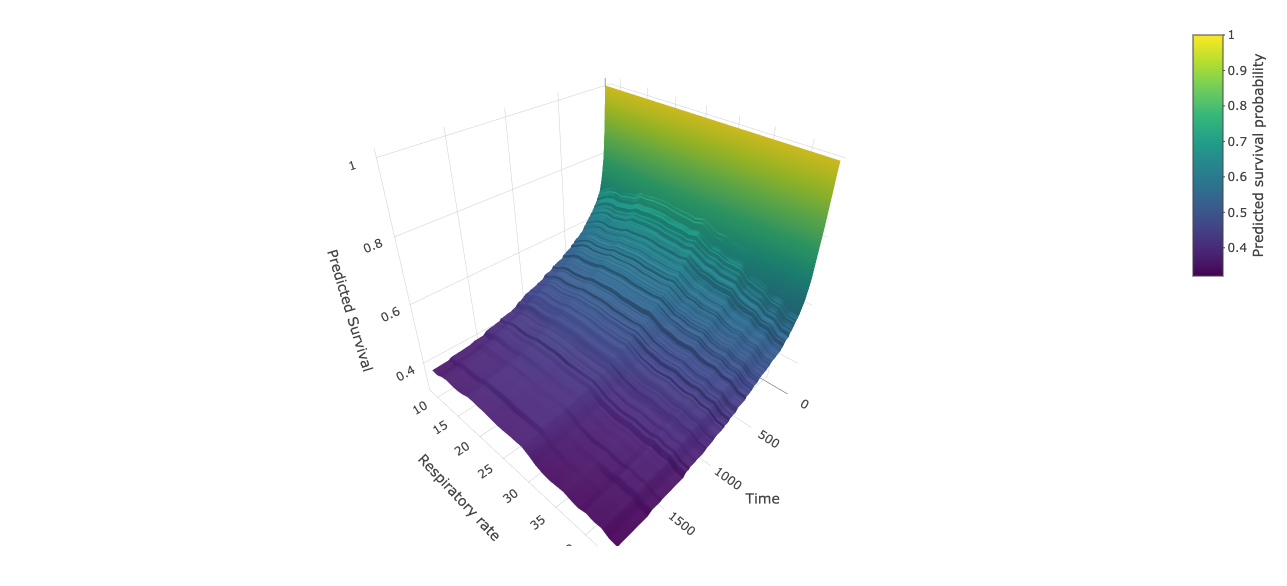}
	\caption{2D plot (left) and 3D plot (right) generated by the survivalContour R package
 %for the Veterans' Administration Lung Cancer study data 
 for SUPPORT III study
 using random survival forests model.}
	\label{rf}
 \end{subfigure}
 \begin{subfigure}[c]{\textwidth}
	\includegraphics[width=0.49\textwidth]{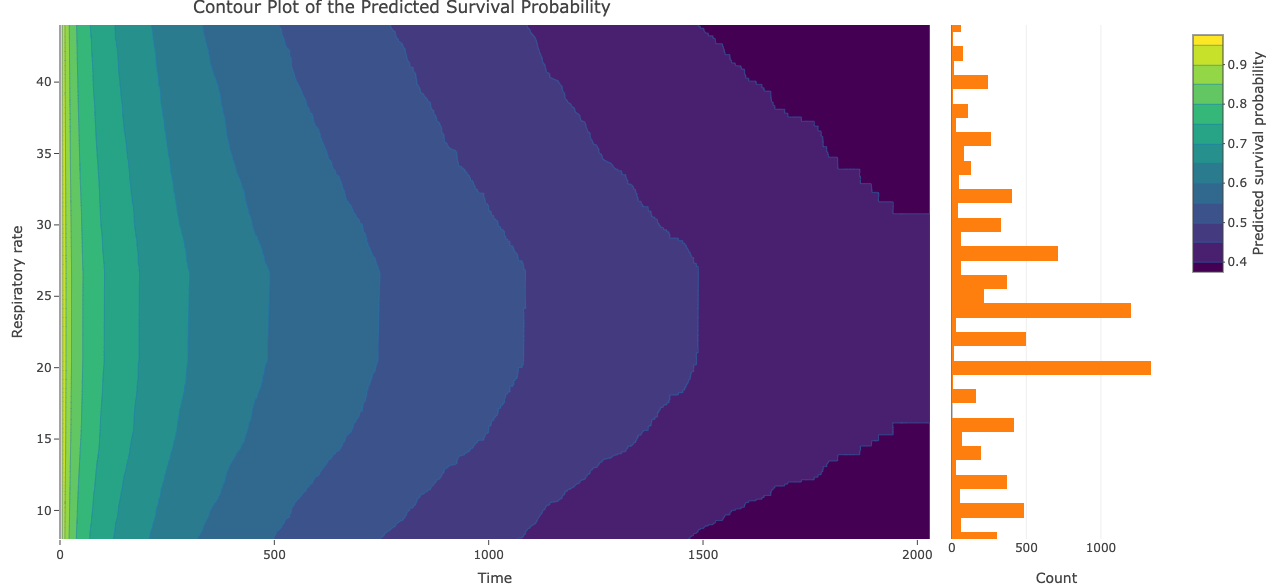}
 	\includegraphics[width=0.49\textwidth]{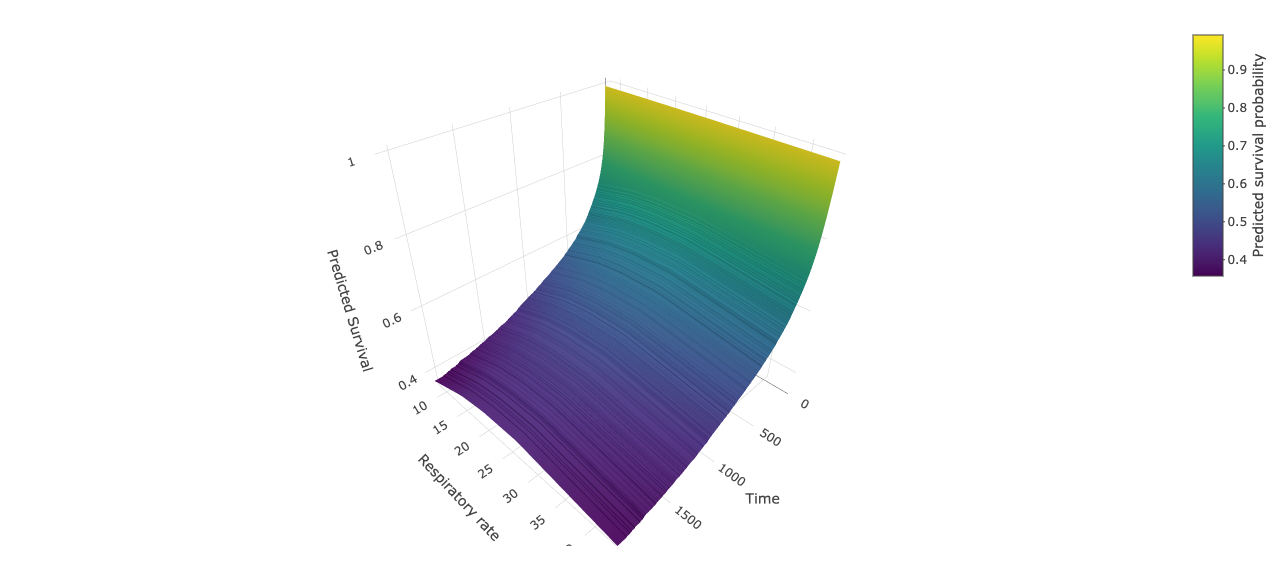}
	\caption{2D plot (left) and 3D plot (right) generated by the survivalContour R package to depict predictions from a DeepSurv model.}
	\label{support}
\end{subfigure}
\begin{subfigure}[c]{\textwidth}
  	\includegraphics[width=0.475\textwidth]{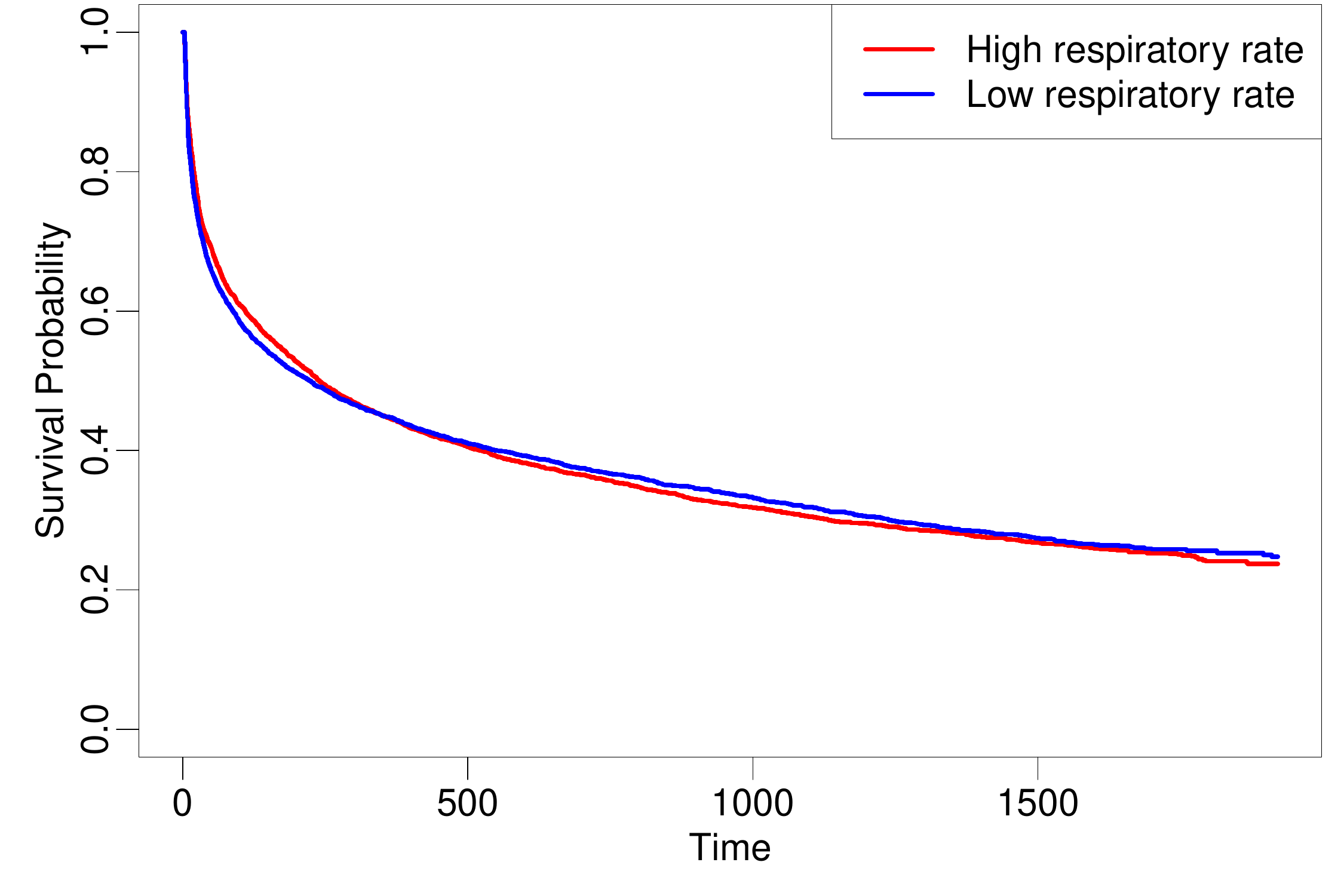}
	\includegraphics[width=0.475\textwidth]{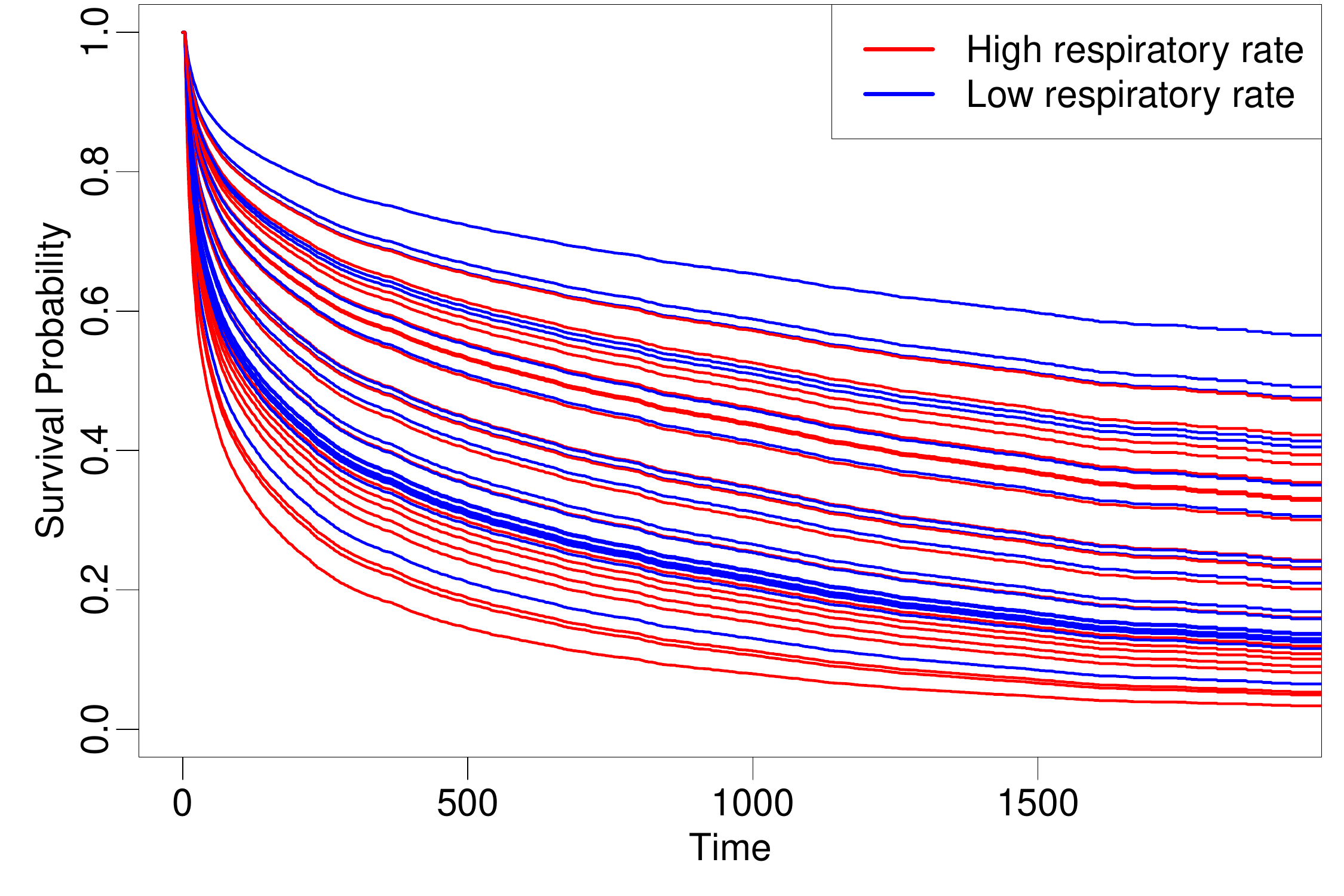}
	\caption{Kaplan Meier curves of the high and the low respiratory rate groups (left) and survival probabilities of randomly selected samples from each group (right).}
	\label{traditional}  
\end{subfigure}
\caption{Survival predictions generated using machine learning methods}
\end{figure}

Figure \ref{rf} depicts 2D and 3D survival contour plots showing the relationship between respiratory rate and predicted survival from a random survival forest model. %A histogram of respiratory rate values is shown to the right of the contour plot in the left panel.
The color variation in the contour plot corresponds to different survival probabilities, with darker colors representing worse predicted survival. The 2D and 3D visualizations from the DeepSurv method are shown in Figure \ref{support}. What static Figures \ref{rf} and \ref{support} cannot reveal is that users can interactively obtain predicted survival rates at specific time points and covariate values by gliding the mouse across the plot. Also, 3D interactive graphics allow users to rotate the 3D surface to view them from multiple angles.

%We present the 3D version of this plot in Figure \ref{support_3D}. This is also an interactive graphic, which allows users to rotate the 3D surface to view it from multiple angles. 
Since the random survival forest and DeepSurv utilize different approaches for prediction, the resulting contour plots are not exactly the same. Nevertheless, the non-monotone relation between respiratory rate and survival is captured by both models: survival rates deteriorate at extreme respiratory rates, whereas the median respiratory rate forecasts better survival, aligning with clinical intuition. This delicate relationship would not be captured by a basic Cox model, which imposes a multiplicative hazards assumption. Moreover, the common practice of dichotomizing the predictor would fail to identify this relationship.

% \begin{figure}[H]\ContinuedFloat
% \begin{subfigure}[c]{\textwidth}
% %	\includegraphics[width=0.8\textwidth]{Fig10b_v2.png}
%  	\includegraphics[width=\textwidth]{cnn2.png}
% 	\caption{3D plot generated by survivalContour to depict predictions from a DeepSurv model.}
% 	\label{support_3D}
% \end{subfigure}

% \end{figure}

To illustrate the limitation of existing methods, Figure \ref{traditional} displays Kaplan-Meier curves for both the high and the low respiratory rate groups, using a cutoff set at the median value. These curves appear perfectly smooth due to the large number of subjects in the patient cohort ($>$ 8000).
Additionally, we randomly selected 20 samples from each group and plotted their predicted survival from the DeepSurv model. The non-linear association evident in Figure \ref{rf} and \ref{support} becomes lost in these presentations, which reflect common practice in presenting survival predictions.

\section{Discussion}
Our survivalContour software opens a new avenue for visualizing the relationship between a continuous predictor and survival outcomes. The Shiny app offers a user-friendly guided experience and can accommodate many popular survival models. The full potential of our plotting approach is revealed when incorporated with state-of-the-art machine learning models through the survivalContour R package. 

Importantly, the survival contour plot focuses on displaying predicted survival and does not aim to provide insight regarding model fit. We recommend that users compute Harrell's Concordance Index  \citep[C-index, ][]{HarrCali82}
and the integrated Brier score \citep{GrafSchm99} to evaluate model fit before sharing results from survivalContour.
Complex deep neural network-based models, like DeepHit or DeepSurv, often require significant manual tuning and the best practices for fitting such complex models are beyond the scope of this paper. Additionally, flexible models may be susceptible to overfitting, so it is always advisable to use an independent dataset for validation.

\subsection*{Data availability statement}
The data presented in Section \ref{sec:cox} were originally published by \cite{PeleGome20}. Restrictions apply to the availability of these data, which were used under license for this study. The Veterans' Administration Lung Cancer data analyzed in Sections \ref{sec:stratified} is available in the  R package \texttt{survival}. The Paquid data analyzed in Section \ref{sec:cmprsk}  is available in the  R package \texttt{riskRegression}. The data from the SUPPORT III study analyzed in Section \ref{sec:ml} is available from the GitHub page \url{https://github.com/havakv/pycox/}.

\subsection*{Acknowledgments}
This work was partially funded by National Institutes of Health grants P30CA016672, SPORE P50CA140388, R01 HL158796, R01 HL124112; CCTS TR000371, Cancer Prevention and Research Institute of Texas RP160693, CCSG P30CA016672 and National Science Foundation grant 2310955.
We would like to thank Jonathan Peled and Keimya Sadeghi of Memorial Sloan Kettering Cancer Center for sharing the MSKCC cohort data from \cite{PeleGome20} presented in Section \ref{sec:cox}.

%\bibliography{survCont} 
%\bibliographystyle{IEEEtran}
%\bibliographystyle{agsm}
%\bibliographystyle{abbrvnat}
%\bibliographystyle{unsrtnat}
\end{document}